\documentclass{article}
\usepackage[utf8]{inputenc}

\usepackage{amsmath}
\usepackage{amsfonts}
\usepackage{amssymb}
\usepackage{makeidx}
\usepackage{graphicx}
\usepackage{epsfig}

\newcommand{\beq}{\begin{equation}} \newcommand{\eeq}{\end{equation}}
\newcommand{\bea}{\begin{eqnarray}} \newcommand{\eea}{\end{eqnarray}}
\newcommand{\bear}{\begin{eqnarray*}} \newcommand{\eear}{\end{eqnarray*}}
\newcommand{\lb}{\label}

\usepackage{hyperref}
\hypersetup{colorlinks=true, linkcolor=blue, citecolor=blue, filecolor=blue, urlcolor=blue,
            pdfproducer={Latex},
            pdfcreator={pdflatex}}

\begin{document}

\title{Why can we observe a plateau even in an out of control epidemic outbreak?
     A SEIR model with the interaction of $n$ distinct populations for Covid-19 in Brazil.%
     \thanks{This work was supported in part by CNPq and FAPERGS, Brazilian funding agencies.}}

\author{M. J. Lazo\thanks{matheuslazo@furg.br} and A. De Cezaro\thanks{adrianocezaro@furg.br}\\
     Institute of Mathematics, Statistics and Physics\\
      Federal University of Rio Grande, Rio Grande, RS, Brazil.}

\maketitle


\begin{abstract}
This manuscript proposes a model of $n$ distinct populations interaction structured SEIR to describe the spread of COVID-19 pandemic diseases. The proposed model has the flexibility to include geographically separated communities as well as taking into account aging population groups and their interactions. We show that distinct assumptions on the dynamics of the proposed model lead to a plateau-like curve of the infected population,  reflecting collected data from large countries like Brazil. Such observations aim to the following conjecture: " Covid-19 diseased diffusion from the capitals to the Brazil interior, as reflected by the proposed model, is responsible for plateau-like reported cases in the country". We present numerical simulations of some scenarios and comparison with reported data from Brazil that corroborate with the aforementioned conclusions.
\end{abstract}

\section{Introduction}

The mathematical modeling of infectious disease has played an important role in the definition of action protocols and public policies to combat diseases aiming at the least social impact. It has been successfully used since the Bernoulli's pioneer work \cite{bernoulli} on infectious diseases. An up-to-date example of the importance of mathematical modeling on public policies for epidemic diseases is the recent change in the UK's strategies regards to the  Covid-19 pandemic \cite{Ferguson, BBC}. An extensive review of the mathematical modeling of infectious diseases can be found in \cite{H2000,hethcote1989three, hays, book} and references therein.

Covid-19 is by now known as a rapidly spread contagious diseases, with a first registered outbreak at the end of December 2019 in Wuhan (China) and that from there spread rapidly around the globe causing mortality without precedents in the recent human history e.g  \cite{WLB2020, He, Nande, Teimouri, Prem, Castilho, Giordano} and references therein. Since up-to-day there is no safe vaccine or treatment, the remaining alternative for the public policies to control the outbreak was the social mitigation, tacked in some places to the extreme of severe lock-downs \cite{He, Nande, Teimouri, Prem, Castilho, Giordano}. The main goal of such a strategy was to "flatness the infection curve" such that the public health service could support the eventual high demand for ICU beds \cite{He, Nande, Teimouri, Prem, Castilho, Giordano}.

Many countries have successfully implemented control strategies for the Covid-19 pandemic outbreak employing massive testing and isolation allied with social distance and population mitigation \cite{He, Nande, Teimouri, Prem, Castilho, Giordano}. Such social measures can be identified by the emergence of a plateau in the infected cases reported on such countries \cite{Prem,Ansumali}. Its success can be measure due to the relatively low mortality rate due to Covid-19 and an early secure economy reopening. Although, Brazil started with some kind of social distance measures since the early stages of the epidemic (around the end of March, with the first confirmed case on February 26, 2020), only recently the number of reported daily mortality (daily new infected) seem to have attained a plateau that, by the way, is in a high-level \cite{plateau,plateau2}. Nevertheless, in some countries the public authorities are pointing for a massive economy reopening, in contradiction of the majority of the epidemiological specialist \cite{Park, Oliveira}.

With one eye on divulged official data, that reveal a plateau-like shape for the reported cases from countries as Brazil \cite{plateau,plateau2,real}, for which several mechanisms, like social distance, massive testing and isolation isolation, were never effectively and  seriously implemented, we shall questioning: \textit{"Why can we observe a plateau even in an out of control epidemic outbreak?".} 

In this context, the main objective of the present work is to analyze how a plateau-like in the infected cases can arouse in an out of control epidemic outbreak and to analyze some scenarios for the Covid-19 epidemic in Brazil. Within this purpose, we take into account the two main tools to study the spread of this disease: a simple mathematical model and population interactions. Actually, we propose and analyze an structured SEIR-like compartmental epidemic model. The SEIR model is a generalization of the well-known SIR compartmental model \cite{H2000, book} that includes the exposed fraction of the population, denote by $E$. The literature on compartmental models and its applications in the COVID-19 outbreak has grown as faster as the pandemic itself, making a complete literature overview almost impossible. Therefore, we shall mention publications that are closely related to this manuscript. Age-structured SEIR models with social mixing were analyzed in \cite{Teimouri, Prem, Castilho, Giordano,Ansumali}, where was analyzed the effectiveness of control measures aimed at reducing the social mixing population in different countries as China, Italy, Uk, US, and Brazil. For a comprehensive overview of related works, see \cite{He, Nande, Teimouri, Prem, Castilho, Giordano,Ansumali} and references therein. In special, in \cite{Ansumali} the authors propose to include dynamics for asymptomatic and pre-symptomatic individuals in a modified SEIR model to explain the existing plateau in Covid-19 reported cases. However, such dynamics need to be calibrated, and it is possible only if a massive testing program is considered. Moreover, these approaches can not be applied to explain how a plateau-like phenomenon arises in an out of control epidemic outbreak.

\paragraph{Main contributions:}  Our main contribution with this manuscript is to provide enough shreds of evidence that a structured SEIR model, where the structure is characterized by $n$ distinct populations in interaction (see~\eqref{SEIR}), can predict a plateau-like shape for infected cases. The distinct populations can represent, for example, geographically separated communities,  different groups (such as the elderly or people with co-morbidity ), etc. As we shall see in the analysis that follows (see Section~\ref{sec:facts}), such plateau is a product of a diffusion-like effect withing distinct populations. Although the derived analysis seems to be simple, it is possible to simulate distinct scenarios, reflecting the Covid-19 infected data ranging from countries where strict vertical isolation was implemented to scenarios with practically no population mitigation was adopted. In particular, we conjecture that the Covid-19 diseased diffusion from the capitals to the Brazil interior, reflected by the proposed model, is responsible for plateau-like reported cases in the country.

The work is structured as follows. In Section~\ref{sec:model}, we present the model. We devote Section~\ref{sec:facts} to share some known facts about the proposed model as well as to analyze the scenarios of the model that predicts a plateau-like shape of reported cases. We also analyze what shall be the real-life scenario that reflects such a situation. As we will see, the plateau-like shape can be associated with the diffusion of the disease within populations (see Subsection~\ref{subsec:plateau}. In Subsection~\ref{subsec:VerticalHorizontal}, we briefly discuss how the proposed model can be use to predict the scenario for vertical an horizontal social isolation.  In Section~\ref{num}, we present numerical simulations of a diversity of scenarios (see Subsection~\ref{subsec:4.1}), for with an extensive discussion is provided. Moreover, we also show how the proposed model can describe reported Brazilian Covid-19 infected cases in Subsection~\ref{subsec:4.2}. In subsection~\ref{subsec:4.3}, we show how vertical social isolation is illustrated by the proposed model.  Finally, Section \ref{sec:C} is dedicated to our conclusions.

\section{The model}\label{sec:model}

In the forthcoming analysis, we shall consider a variation of the well-known epidemiological SIR compartmental model proposed by Kermack and Mckendrick \cite{KM1927}. Actually, our model is a generalization of the classical SEIR model \cite{H2000, book} where we introduce the interaction of $n$ distinct populations. Let $N_i$ be the number of individuals in population $i$ ($i=1,2,...,n$), and $N_T=N_1+N_2+\cdots +N_n$ be the total integrated population. Let $S_i$, $E_i$, $I_i$ and $R_i$ be the fractions, in regard to $N_i$, of population $i$ that are susceptible, exposed, infectiously and recovery, respectively, at time $t$. In the model we consider, the time evolution is given by the following dynamic system for $i=1,2,...,n$:
%
%
\beq
\label{SEIR}
\begin{split}
\dot{S_i}&= - S_i\sum_{j=1}^n \beta_{ij}I_j + \mu_i(1-S_i)\\
\dot{E_i}&= S_i\sum_{j=1}^n \beta_{ij}I_j - (\alpha_i+\mu_i)E_i\\
\dot{I_i}&= \alpha_iE_i - (\gamma_i+\mu_i)I_i\\
\dot{R_i}&= \gamma_iI_i-\mu_iR_i,
\end{split}
\eeq
where $\beta_{ij}$ are the disease transmission rate (proportional to the average contact rate in the population and within the population), $\alpha_i$ is the inverse of the incubation period, $\gamma_i$ is the inverse of the mean infectious period, for $i=1\cdots, n$. Furthermore, we assume that mortality rates $\mu_i$ is equal to the birth rates, such that the total population $N_i$ is constant during the diseases. Hence, we have $S_i(t)+ E_i(t)+ I_i(t) + R_i(t) = 1$. 

Moreover, the dynamics~\eqref{SEIR} is considered with the following initial conditions
\beq
\lb{IC-SEIR}
S_i(0)= 1-I_i(0), E_i(0)=0, I_i(0), R_i(0) = 0\,,
\eeq
where $I_i(0)$ is the fraction of the population $i$ infected at the time $0$.

\section{Some facts about the model}\label{sec:facts}

In this section, for easiness, we shall consider the model~\eqref{SEIR}-\eqref{IC-SEIR} with $\mu_i = 0$. Notice first that, from the third equation in~\eqref{SEIR}, we get that the variation of $I_i(t)$ is proportional to the variation of $E_i(t)$. In this setting, summing up the second and the third equation in \eqref{SEIR}, we have the well-known SIR model, with $\hat{I}_i(t) = I_i(t) + E_i(t)$. We consider that the properties we shall explore will be well understood if some well-known facts about the model~\eqref{SEIR}-\eqref{IC-SEIR} are revisited. We provide some details here for the sake of completeness. 
The model~\eqref{SEIR}-\eqref{IC-SEIR} has a unique smooth solution, that depends continuously on the parameters and the initial conditions, e.g. \cite{Sotomayor}.
Then, whenever $S_i(0)$ and $\hat{I}_i(0)$ are strictly positive, results in $S_i(t)$ and $\hat{I}_i(t)$ non negative, for all $i=1, \cdots, n$.  It follows from the first equation in~\eqref{SEIR} that $S_i(t)$ is decreasing. Moreover, adding the first three equations in~\eqref{SEIR}, we get the following conservation law
\beq \label{eq:conservation}
S_i(t) + \hat{I}_i(t) + \gamma_i \int_0^t \hat{I}_i(s) ds = S_i(0) + \hat{I}_i(0)\,. 
\eeq

From~\eqref{eq:conservation}, we get that $\hat{I}_i(t) < \infty$, for all $t\geq 0$. Furthermore, $\int_0^\infty \hat{I}_i(t) dt \leq \infty$. Since, $S_i(t)$ is decreasing, \eqref{eq:conservation} also implies that $S_i^\infty:=\lim\limits_{t \to \infty}S_i(t) > 0$. Consequently, \eqref{eq:conservation} also imply that $\lim\limits_{t \to \infty} \hat{I}_i(t) =0$. 
The \textit{basic reproduction number} of the population $i$ ,$\mathcal{R}_0^i$, is the quantity that express the expected number of cases directly generated by one case in a population and withing the selected population, at the initial phase of the infection, is defined as $\frac{\beta_{ii}S_i(0)}{\gamma_i}$. It is well-known that if $\mathcal{R}_0^i > 1$, $\hat{I}_i(t)$ start increasing and then decreasing. Therefore, given the smoothness of $\hat{I}_i(t)$ (as a result of the existence of a solution of~\eqref{SEIR}) and the above mentioned properties of $\hat{I}_i(t)$, we conclude that it trajectory has a concave hump with extremes in $\hat{I}_i(0)$ and $\hat{I}_i(\infty) = 0$, see Figure~\ref{fig1}. Hence, $\hat{I}_i(t)$ attains a maximum at a point $t_p^i$, known as the turning point. Adding the second and third equation in~\eqref{SEIR} and the fact that $\dot{\hat{I}}_i(t_p^i) = 0$, implies that 
\beq \label{eq:Smax}
S_i(t_p^i) = \frac{\gamma_i}{\beta_{ii}}\left(\frac{1}{1 + \sum\limits_{j \neq i}^n \frac{\beta_{ij}}{\beta_{ii}} \hat{I}_j(t_p^i) }\right)\,,\quad \mbox{for } i \in \{1,\cdots, n\}\,.
\eeq    

Although simple, the analysis of the equation~\eqref{eq:Smax} revels several scenarios for the dynamics, as pointed below:
\begin{enumerate}
    \item Assuming that we have $n$ isolated populations, i.e., $\beta_{ij} = 0$ for $j \neq i$, then $$S_i(t_p^i) = \frac{\gamma_i}{\beta_{ii}}\,,$$
    as in the canonical SIR model. In this particular setting of no-cross contamination between different populations, we have $t_p^i = t_p^j$ and then, we see the pick of the populations as one. 
    
    \item It is clear from~\eqref{eq:Smax} that $S_i(t_p^i)$ is lower and therefore $\hat{I}_i(t_p^i)$ is higher, depending on the infected connection network, since the quantity $\sum\limits_{j \neq i}^n \frac{\beta_{ij}}{\beta_{ii}} \hat{I}_j(t_p^i)$ appear in the quotient. It implies that the infection will be higher as higher is the infected  connection network. Consequently, any kind of population isolation shall be tacked horizontally on the populations, in order of the diseases spread mitigation be effective.   
    
    \item Let $\mathcal{I}(t): = \sum_{i=1}^n \hat{I}_i(t)/n$ and $\mathcal{S}(t) : = \sum_{i=1}^n S_i(t)/n$ the total proportion of the population infected/exposed  and susceptible, respectively. By adding~\eqref{eq:conservation}, we find out that the same behaviour discussed above is true also for $\mathcal{I}(t)$. Indeed, its expected trajectory is a positive concave hump, beginning at $\mathcal{I}(0)$ and ending at $\mathcal{I}(\infty)$. Furthermore, the maximum of $\mathcal{I}(t)$ will occurs at $t_p^{i^*}$ (that is population dependent) related to maximum value of $\sum\limits_{j \neq i}^n \frac{\beta_{ij}}{\beta_{ii}} \hat{I}_j(t_p^{i^*})$.
    
\end{enumerate}

\subsection{The emerging plateau-like infected curve: A consequence of diseases diffusion within distinct populations}\label{subsec:plateau}

A more interesting question that we would like to discuss in this approach is the observed plateau-like phenomena  as a consequence of the diseases diffusion. Below, we shall argue how the proposed model~\eqref{SEIR} can be used to describe such phenomena in the exposed/infected dynamics. 

It started as follows: Assume that the diseases started at the population $i=1$, i.e., at time $t=0$, we have $\hat{I}_1(0)> 0$ and $\hat{I}_i(0)=0$, for $i>1$. Hence, the diseases will spread out throughout the population thanks to the population interaction (see the first equation in~\eqref{SEIR}). Therefore, it would take some time for the diseases spread out to the entirely population.  It turns out that, for any turning point $t_p^i$, the quantity $\sum\limits_{j \neq i}^n \frac{\beta_{ij}}{\beta_{ii}} \hat{I}_j(t_p^i)$ will present a sort of equilibrium (as $i$  changes) since for the population $j$, with $j$ relatively far-way form $i$, the infection is starting to growth and them $\hat{I}_j(t_p^i)$ have a relatively small value, (for example, is we are looking for $i$ close to $1$ and $j$ close to $n$), or then, the opposite situation holds; the infected proportion of the population already start declining, consequently $\hat{I}_j(t_p^i)$ is relatively small ( for example, is we are looking for $i$ close to $n$ and $j$ close to $1$). As far as  $\beta_{ij}/\beta_{ii}$ remains proportion throughout the population, the diffusion phenomena withing populations takes $S_i(t_p^i)$ at the same level (see \eqref{eq:Smax}) . As a consequence, the picks $\hat{I}_i(t_p^i)$ forms a plateau-like effect observed in the infected/exposed compartment $\mathcal{I}$. By the way, this phenomena is observed for the COVID-19 real data on large territorial counties as US and Brazil, see Figure~\ref{fig4}.

Another consequence of the diffusion phenomena discussed above is that the largest $n$ is longer will take for the diseases starts to decline. Consequently, the plateau-like phenomena is larges (see Figure~\ref{fig2}). Moreover, it follows from~\eqref{eq:Smax} that it takes higher levels as higher are the proportion of $\beta_{ij}/\beta_{ii}$, as expected in largest populated cities or metropolitan regions. 

\subsection{ Vertical and horizontal social isolation: How far we can get into this discussion with the SEIR model~\eqref{SEIR}} \label{subsec:VerticalHorizontal}

Since the COVID-19 pandemic outbreak and the lack of a treatment with efficacy evidences, the social isolation is the unique known measure for the pandemic controlling, e. g. \cite{Ferguson, BBC, WLB2020, He, Nande, Teimouri, Prem, Castilho, Giordano} and references therein.  Some counties as UK and Sweden pointed or adopted a vertical social isolation strategy (to protect only  elderly or  co-morbidity proportional of the population), contrarily to the World Health Organization recommendation of a horizontal social isolation. Regardless of economical impacts of the adopted social isolation strategy, in this section we discuss how far the proposed model~\eqref{SEIR} can share light on the best strategy to be adopt to protect the population. It is worth to mention that we do not intend to exhaust the discussion on the subject, but only to make some comments on the subject based on the model~\eqref{SEIR}.

At this point is reasonable to assume an age-structured population, where $i=1$ represents the proportion of elderly and co-morbidity population. Although the vertical isolation seem reasonable (it is equivalent to take $\beta_{11}$ as lower as possible, but of course, it is impossible assume it to be zero), since the mortality for the COVID-19 are concentrated mostly in the elderly/co-morbidity proportion of the population \cite{WHO}, it follows from~\eqref{eq:Smax} that $\beta_{11}$ lower will get effect on the number of infected at the class $i=1$, at least on it pick, might be compensate for the interactions with the other ages-classes ($\beta{ij} \neq 0$) on the day-to-day life. The particular in-housing interaction between many aging people are the reality of many families on lower incoming countries like Brazil \cite{WHO}. Notice that is sufficient that $\hat{I}_j(t_p^i)$, for one $j\neq 1$ with a high value to take $S_1(t_p^i)$ down (and consequently $\hat{I}_1(t_p^i)$ be large). See the scenario presented in~\ref{subsec:4.3} for a numerical representation of such situation.    

On the other hand, in a horizontal isolation scenario we have $\beta_{ij}$ (consequently, $\hat{I}_j(t)$ for all $t$) closed to zero, for all $i,j\in\{1,\cdots, n\}$. Hence, it follows from~\eqref{eq:Smax} that quantity $S_i(t_p^i)$ is large resulting in $\hat{I}_i(t_p^i)$ relatively closed to $\hat{I}_i(0)$, meaning that pressure on the health system might be kipped under control.

\section{Numerical solutions}\lb{num}

In this section we present numerical solutions for some particular cases of our model \eqref{SEIR}. We consider, for simplicity, that all the $n$ populations have the same number $N$ of individuals, namely, $N_i=N=1000000$. We also consider the simplest interaction case where each population interacts with only two neighbour populations, namely, $\beta_{ij}=\overline{\beta}\neq 0$ if $j=i\pm 1$, $\beta_{ij}=0$ for $|j-i|>1$. In this case we have a one dimensional chain of populations where $|j-i|$ can be used to define a distance from populations $i$ and $j$. Furthermore, in order to avoid asymmetries in the interactions of the first ($i=1$) and the last ($i=n$) populations, we also consider $\beta_{1n}=\beta_{n1}=\overline{\beta}$ (this consideration implies that we have a periodic uni-dimensional chain).
Finally, we set $\beta_{ii}=\beta$, $\alpha_i=\alpha$, $\gamma_i=\gamma$, $\mu_i=\mu=0$ (homogeneous populations). All numerical solution are obtained by the traditional backward finite difference method with a time step of $\Delta t=0.5$ days. 

\subsection{Several distinct populations}\label{subsec:4.1}

We consider in this case $\alpha=1/5$ (we are supposing that the meantime before an exposed individual becomes infectious is five days), $\gamma=1$ (the meantime before isolation for infectiously individuals is one day),  $\beta=1.5$ to result in a severe epidemic outbreak, and $\mu=0$. For the interaction between neighboring populations we consider $\overline{\beta}=0.02$, that is the contagious rate from individuals outside the population is seventy-five times smaller than from individuals of the same population. Initially, we consider that there is $200$ infectious individual in population $i=1$ and zero in the other populations.

\begin{figure}[h]
\includegraphics[width=0.49\textwidth]{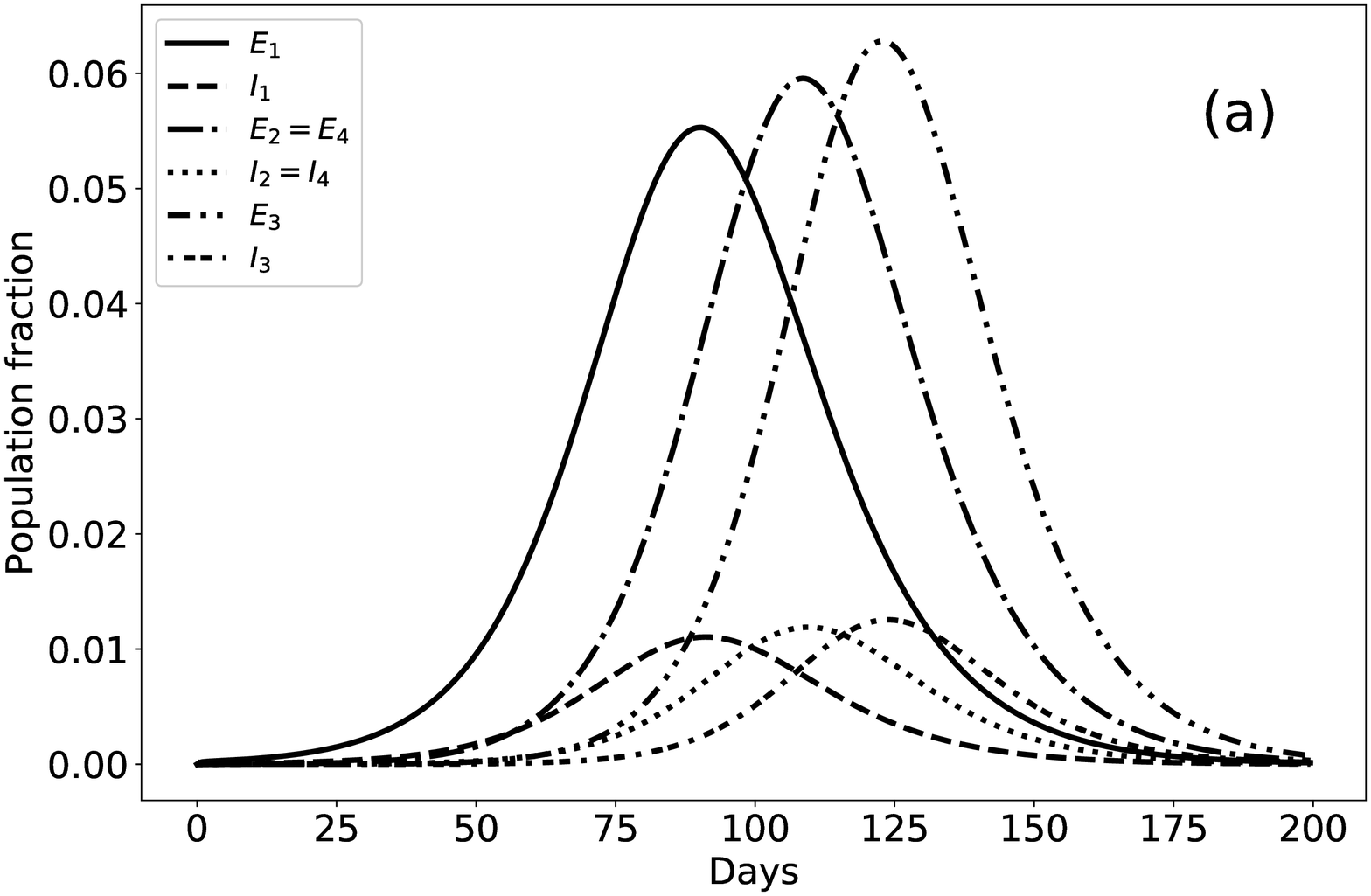}
\includegraphics[width=0.49\textwidth]{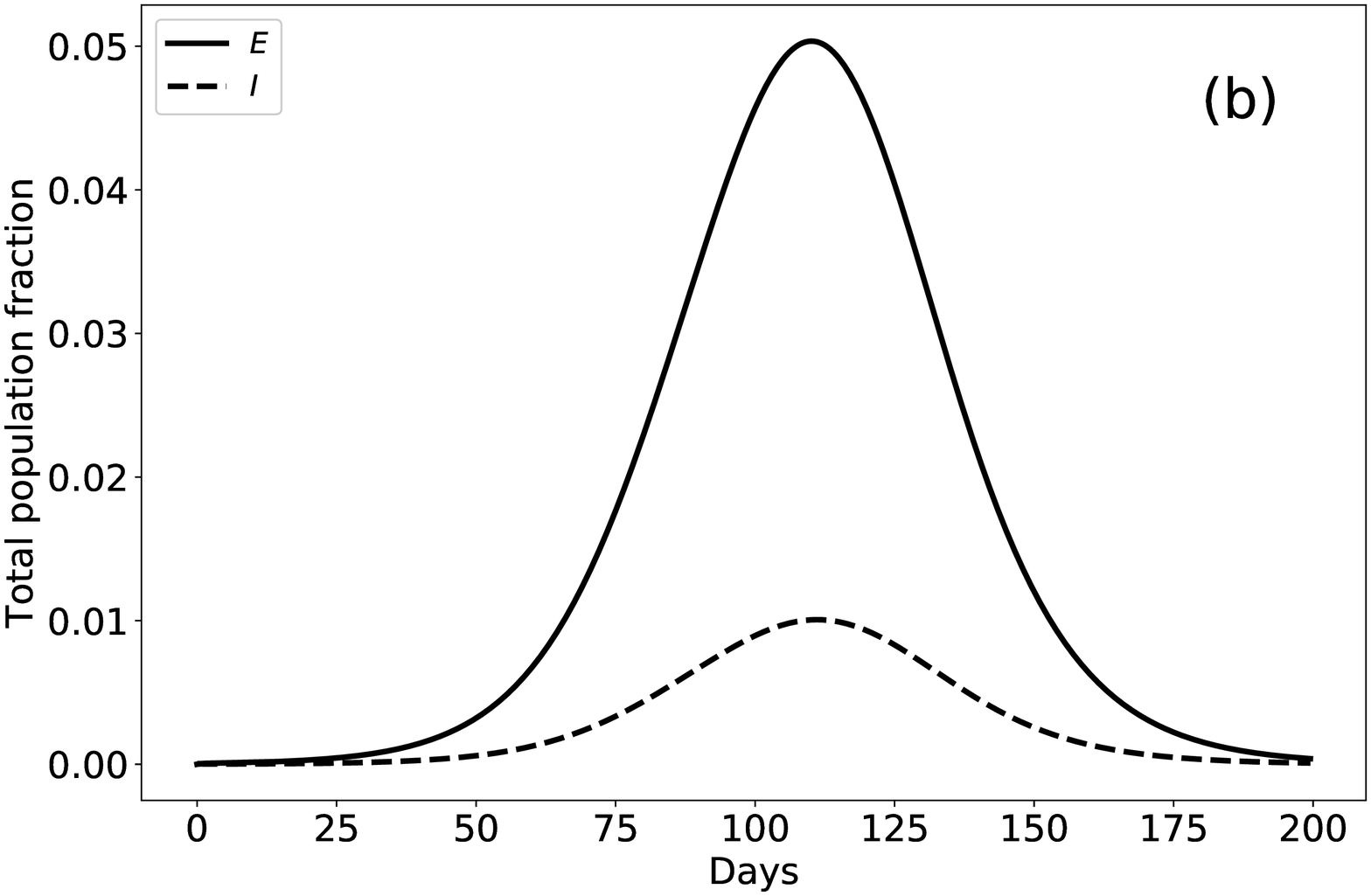}
\caption{$n=4$, $\beta=1.5$, $\overline{\beta}=0.02$ \label{fig1}.}
\end{figure}

In Fig. \ref{fig1} we show the role of interaction $\overline{\beta}$ in each population for $n=4$. In Fig. \ref{fig1} (a) we show the time evolution of exposed $E_i$ and infected $I_i$ compartments. The exposed and infected compartments of population $i=1$ reaches its maximum point first, followed by populations $i=2$ and $i=4$. The last population to reach its epidemic peak is $i=3$. These results are consistent since the epidemic outbreak starts in population $i=1$, propagating to the first neighbouring populations $i=2$ and $i=4$, and finally achieving the most distant population $i=3$. 
The total exposed $E=(E_1+E_2+E_3+E_4)/4$ and infected $I=(I_1+I_2+I_3+I_4)/4$ fractions in relation to the total population $N_T=N_1+N_2+N_3+N_4$ is displayed in Fig.\ref{fig1} (b). 

\begin{figure}[h!]
\includegraphics[width=0.49\textwidth]{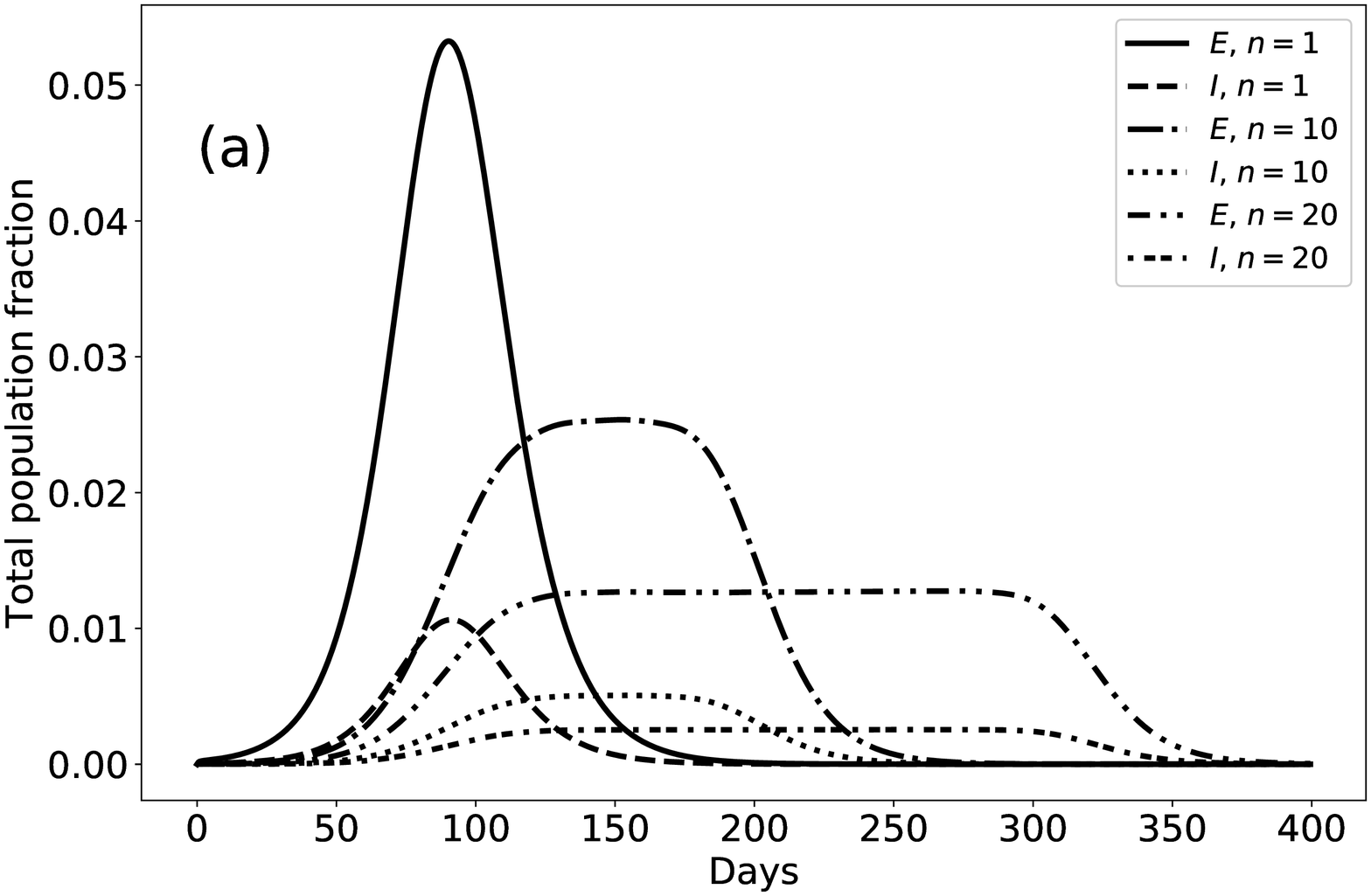}
\includegraphics[width=0.49\textwidth]{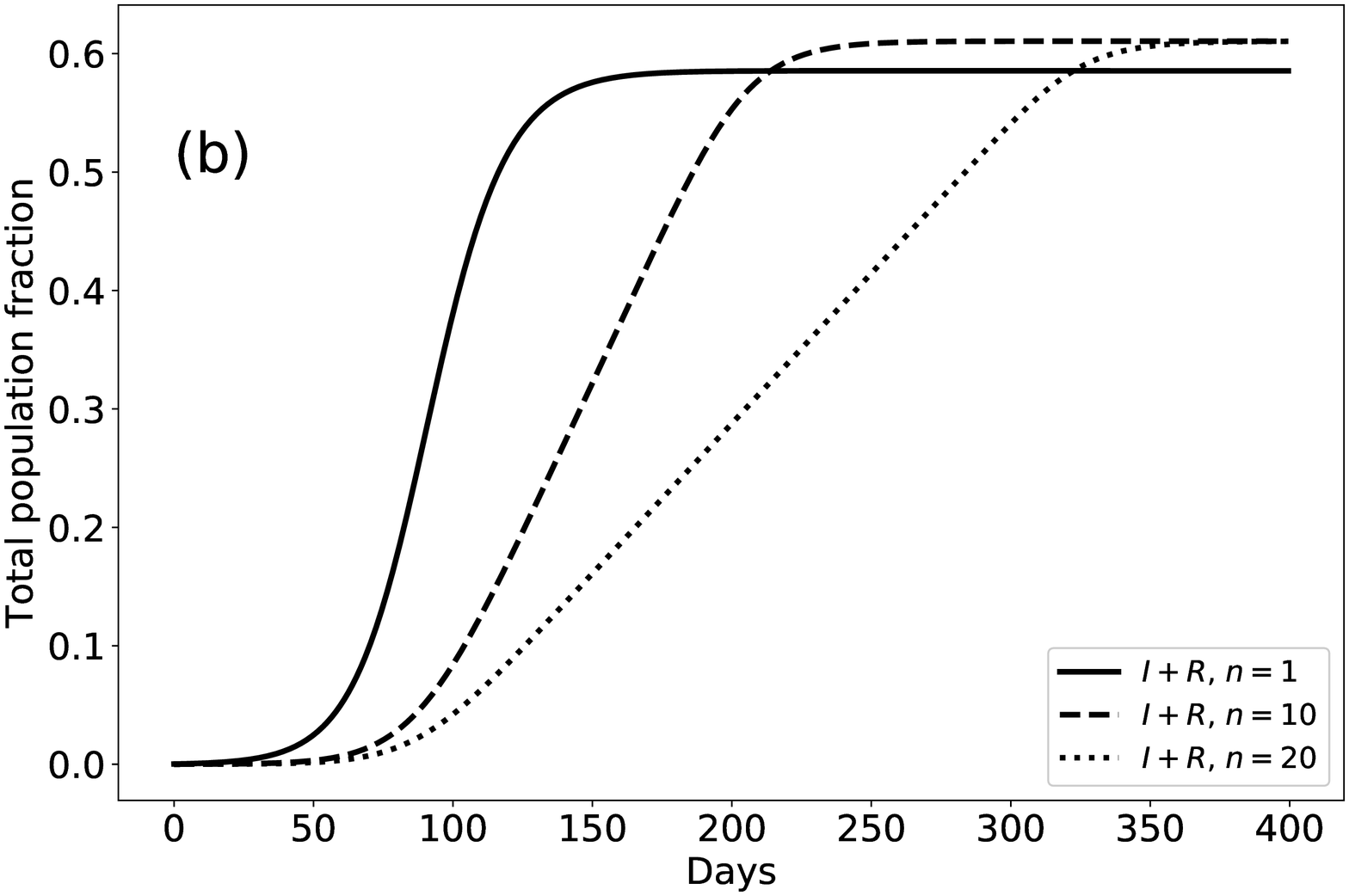}
\caption{$\beta=1.5$, $\overline{\beta}=0.02$ \label{fig2}.}
\end{figure}

The difference in the time that each population reach its maximum point have as a consequence the emergency of a plateau when the number $n$ increase. In Fig. \ref{fig2} (a) we show the total exposed $E=\sum_{i=1}^nE_i/n$ and infected $I=\sum_{i=1}^nI_i/n$ fractions for $n=1,10,20$, the case $n=4$ was displayed in Fig. \ref{fig1} (b). For $n=1$ we have the classical SEIR model \cite{H2000, book} with a sharp curve for both exposed and infected fractions. For $n=4$ (see Fig. \ref{fig1} (b)), both curves broaden but still there are not a plateau. By increasing the number of distinct populations to $n=10$ a well-defined plateau arouses. For $n=20$ we can see that the plateau extension increases with the growth of $n$. As a consequence of the appearance of the plateau, we have an increase in the duration of the epidemic outbreak, but with a smaller fraction of sick individuals in the population at every moment. However, the accumulated fraction of sick people ($ I + R $) at the end of the outbreak varies little with $ n $, as can be seen in Fig. \ref{fig2} (b).

\begin{figure}[h!]
\includegraphics[width=0.98\textwidth]{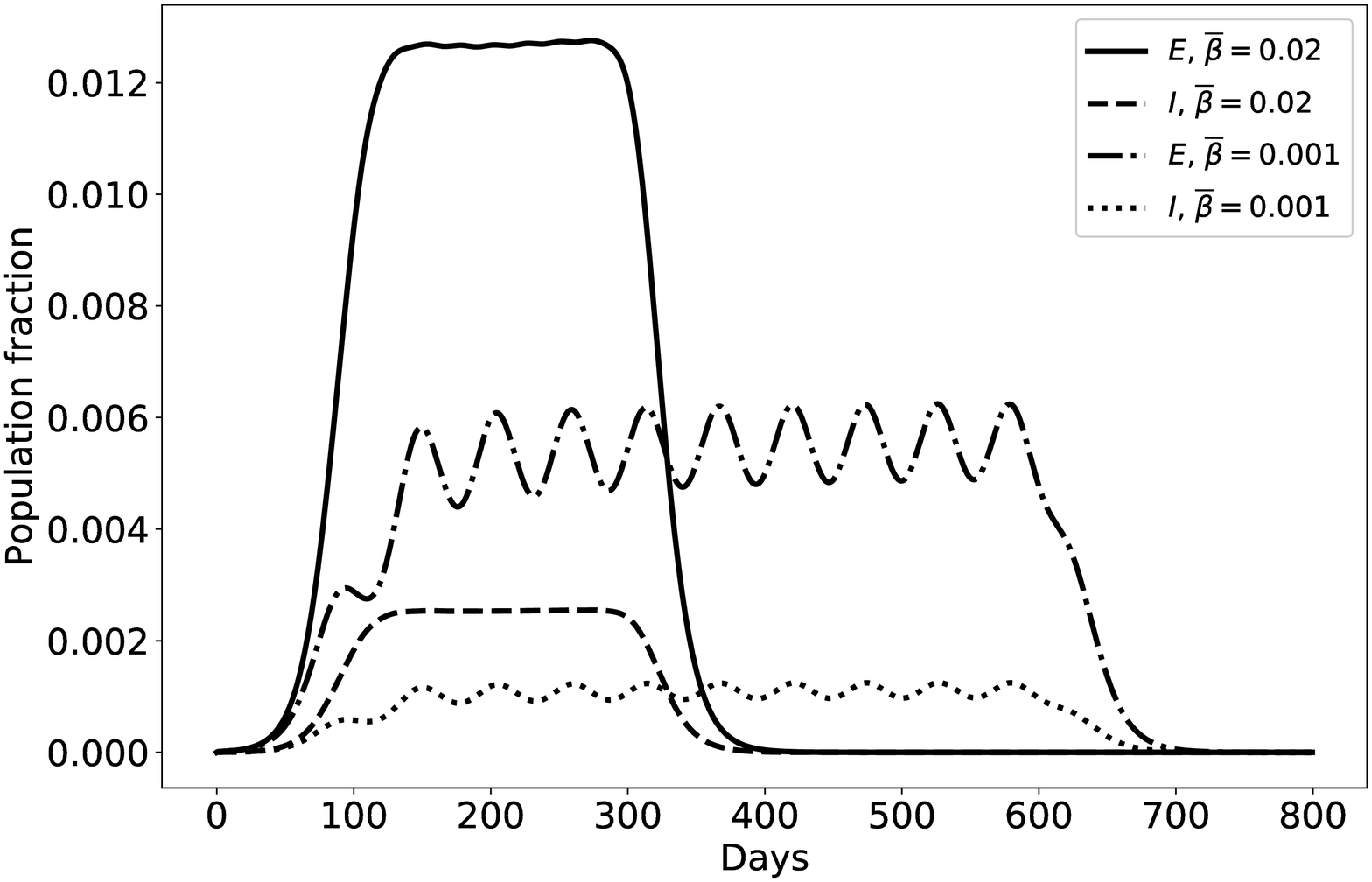}
\caption{$n=200$, $\beta=1.5$, $\overline{\beta}=0.02$ \label{fig3}.}
\end{figure}

Finally, Fig. \ref{fig3} displays the role of the intensity of interaction $\overline{\beta}$ for the epidemic outbreak dynamics. We consider two scenarios with $n=20$. For $\overline{\beta}=0.02$ we have a strong interaction between distinct populations (a population is heavily contaminated by neighboring populations). As a consequence, we observe a plateau that results from small-time distances between the epidemic peaks for each individual population $i$. For a small interaction $\overline{\beta}=0.001$ (distinct populations are almost isolated) the time-distances between the epidemic maximum points for each individual population $i$ increases, resulting in an oscillatory behavior for exposed and infected fractions (instead of a plateau) and in a long total time duration for the epidemic outbreak. 

\subsection{Comparing with real data from Brazil}\label{subsec:4.2}

\begin{figure}[h!]
\includegraphics[width=0.49\textwidth]{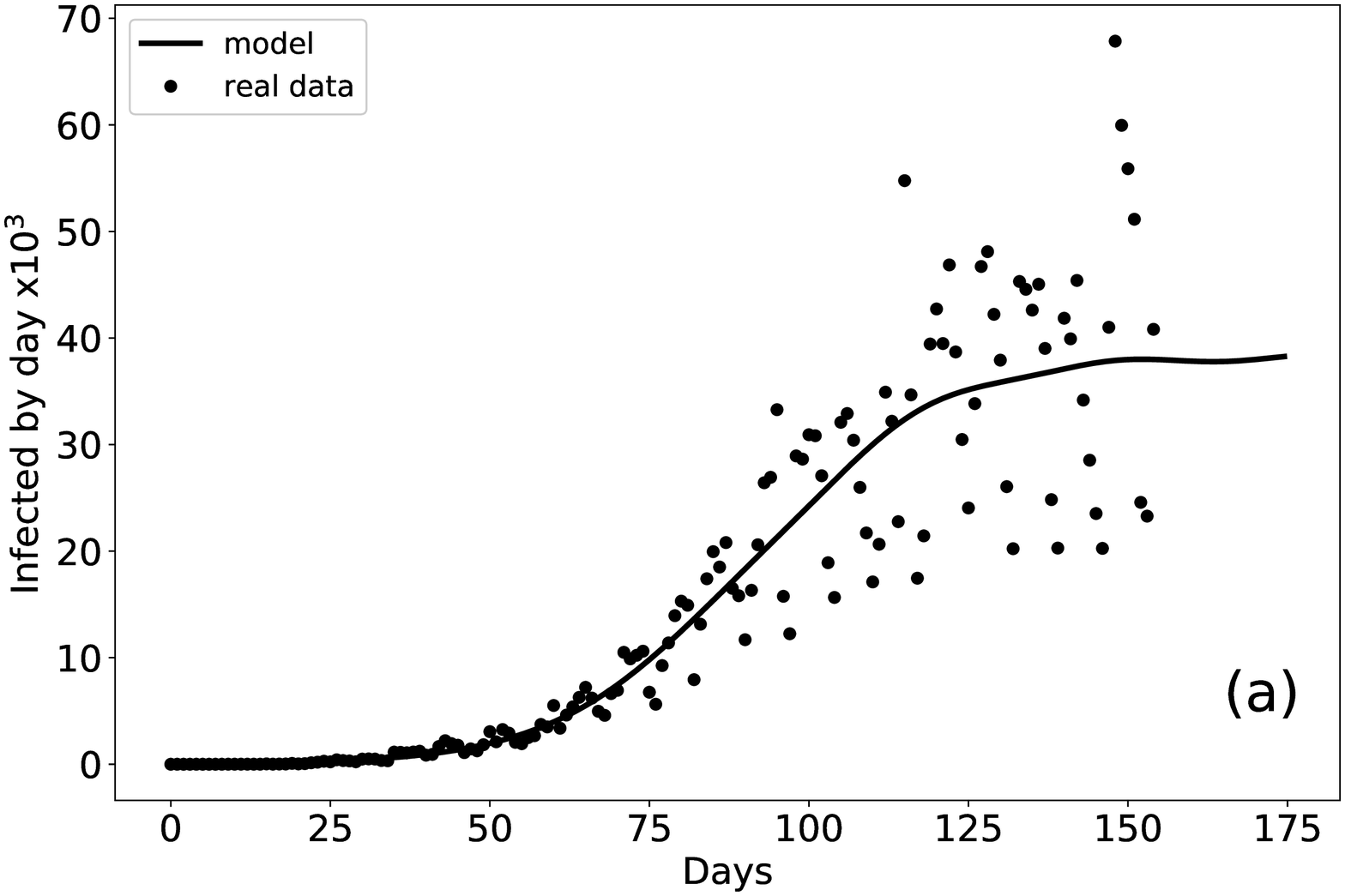}
\includegraphics[width=0.49\textwidth]{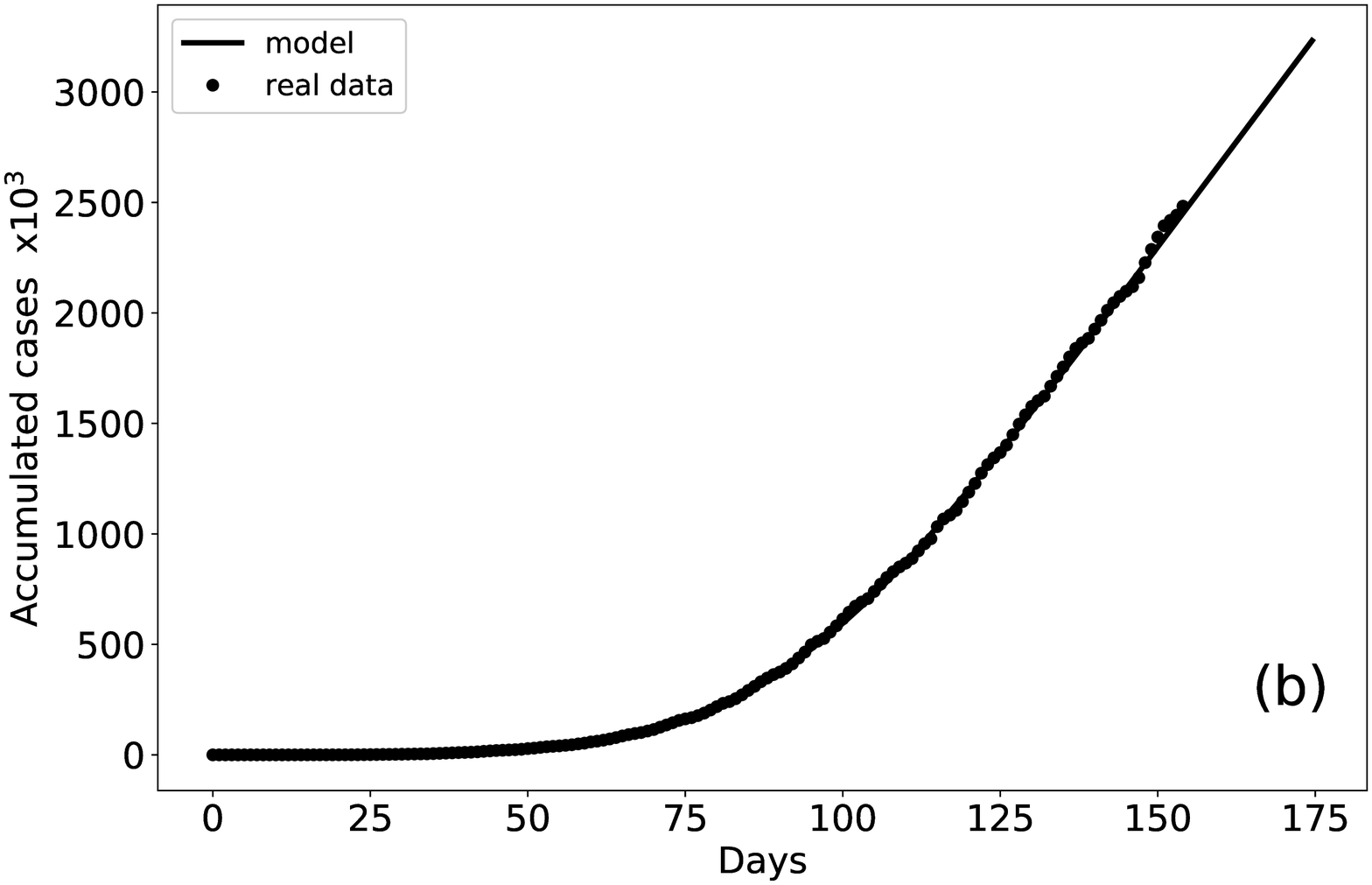}
\caption{$n=200$, $\beta=1.51$, $\overline{\beta}=0.01$, $\gamma=1$, $\alpha=1/5.2$ \label{fig4}.}
\end{figure}

It is important to stress that it is not the objective of the present work to describes real data by using realistic values for the model parameters (as realistic values for populations $N_i$ and parameters $\alpha_i$, $\beta_{ij}$, $\gamma_i$ and $\mu_i$). The result displayed here shows that our model has a great potential to describe real epidemic outbreak, but a detailed stud of real data will be done in a future work. Our present objective is only to show that the model has the ingredients to correctly describes the behavior of Covid-19 dynamics found in real data from Brazil, including the appearance of a plateau phenomenon. With $n=200$ and $N_i=N=1000000$ we found that our model describes very well the behavior of Covid-19 epidemic outbreak when compared against real data from Brazil (see Fig. \ref{fig4}). The real data displayed in Fig. \ref{fig4} is from the ministry of health of Brazil from February 26 up to July 28 of 2020 \cite{real}. The model parameters are set to $\alpha=1/5$, $\beta=1.5$, $\overline{\beta}=0.01$ and $\gamma=1.004$. Fig. \ref{fig4} (a) shows the diary number of infected individual, and Fig. \ref{fig4} (b) displays the accumulated number of sick people ($ I + R $). The emergence of a plateau in the real data is characterized by an almost constant mean value for new reported daily cases, which results in an almost linear growth for the total accumulated cases. By looking at the real data in Brazil, in Figures \ref{fig4} (a) and (b), we can stipulate that the Covid-19 epidemic outbreak reaches a plateau in Brazil approximately four months after the first reported case (at February 26). From these figures, we see a good concordance between real data and our model both in the emergence of a plateau and in the general curve shape of the daily and total accumulated cases, respectively. This result indicates that the plateau of the Covid-19 epidemic outbreak in Brazil may be a consequence of the disease's geographic diffusion. This fact motivates further research aiming at a more accurate description of the real problem through a more realistic choice for our model's parameters.

\subsection{Vertical social isolation}\label{subsec:4.3}

Finally, in addition to populations in different geographic regions, our model \eqref{SEIR} can also be used to describes different partitions of a population in the same geographic location. We consider as an example the problem of vertical social isolation. In this case, a number of $N_1$ individuals in risk groups (such as the elderly or people with co-morbidity) are isolated, while the rest $N_2$ individuals of the population are not confined. Let us assume that isolated individuals are confined alone. In this case, we should have $\beta_{11}=0$ since isolated persons do not have contact with other isolated individuals. The disease contagion follows from contact with non-confined persons. Let also we set $\beta_{22}=1.5$ for the disease transmission rate of non-isolated individuals (proportional to the average contact rate in the $i=2$ population). By supposing a hypothetically simple case where non-confined persons go out from home five times a week, while isolated individuals go out only one time a week, we can set $\beta_{12}=\beta_{21}=\beta_{22}/5=0.3$. 

\begin{figure}[h!]
\includegraphics[width=0.49\textwidth]{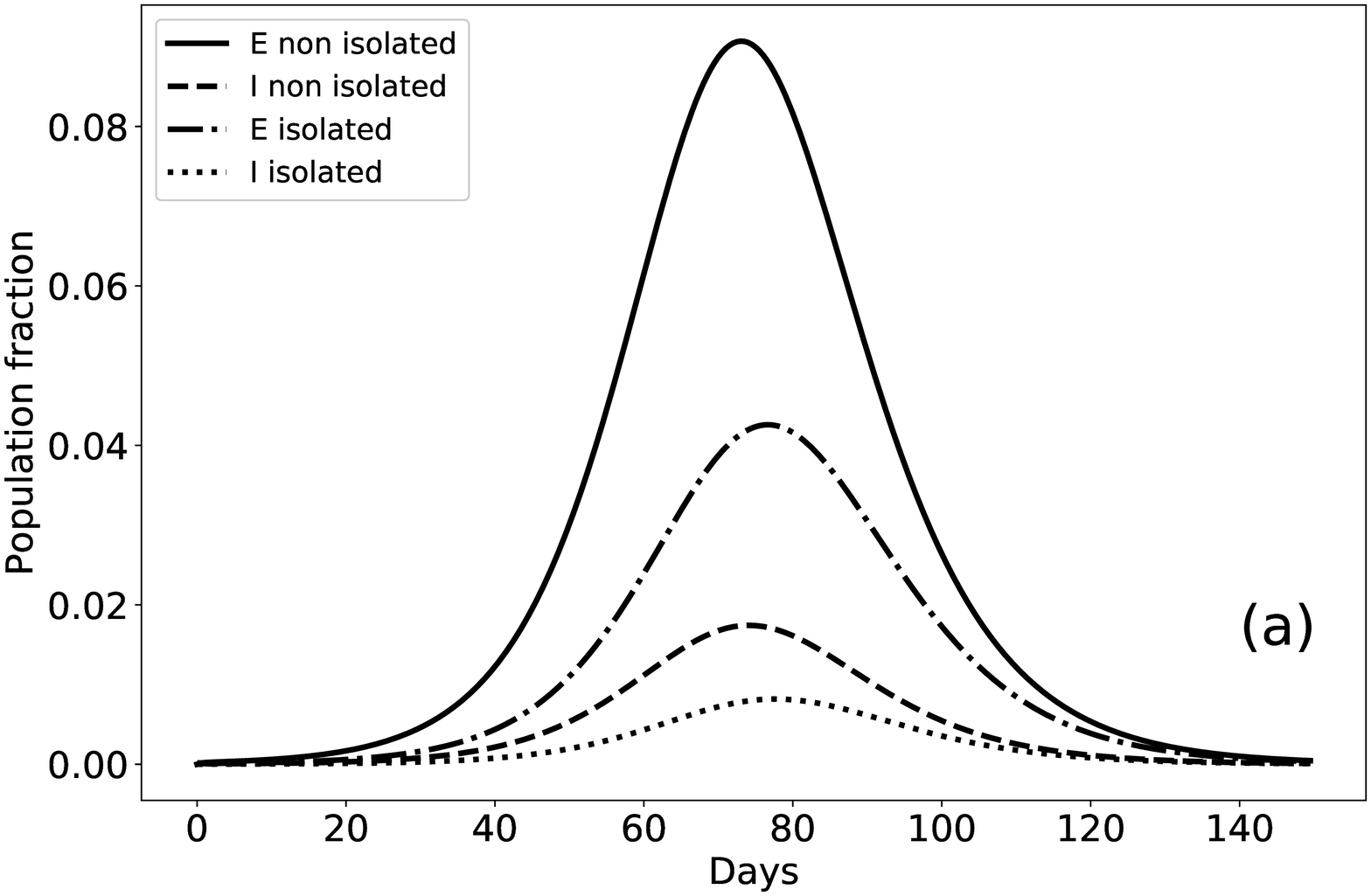}
\includegraphics[width=0.49\textwidth]{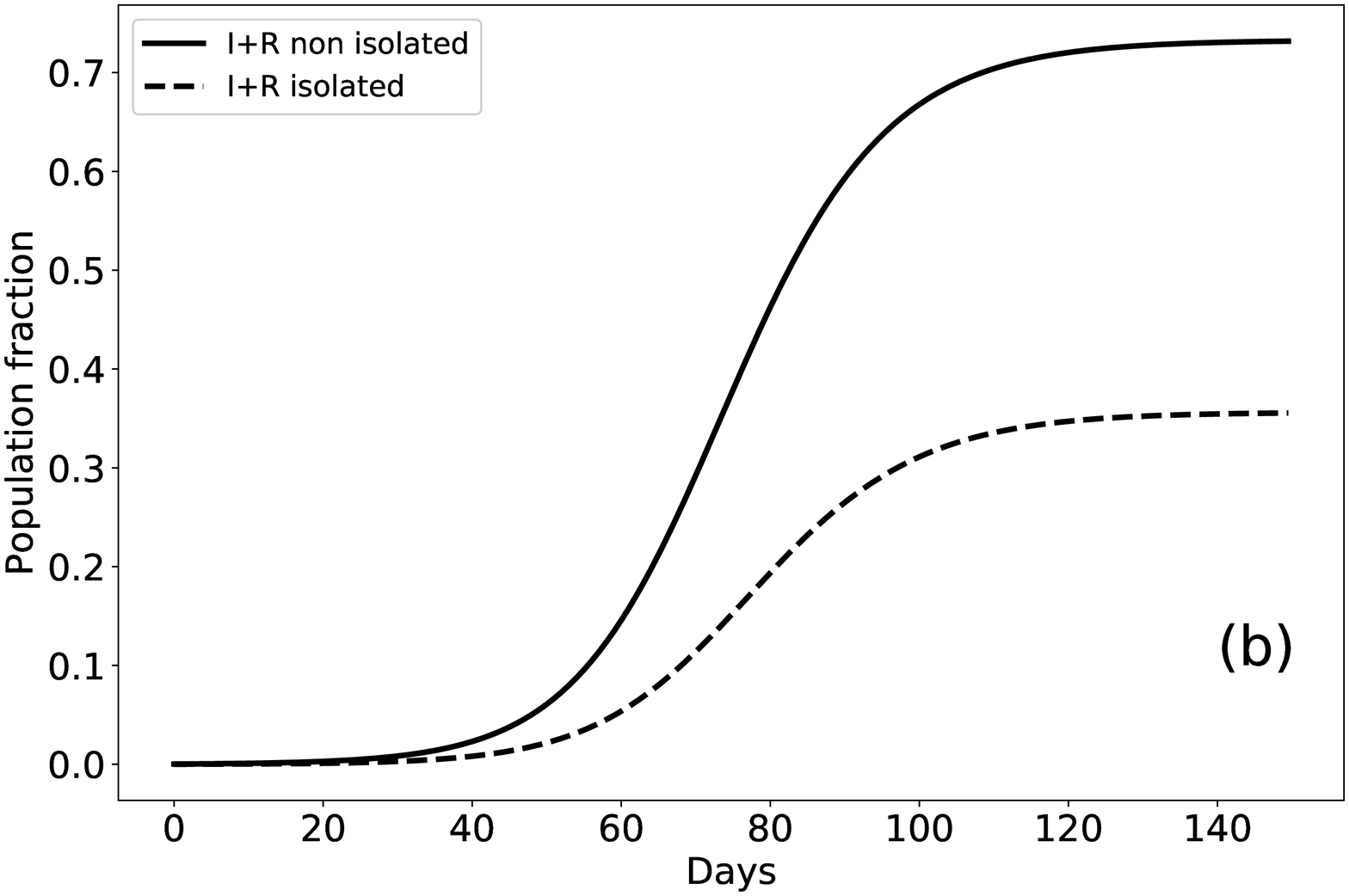}
\caption{$n=2$, vertical isolation \label{fig5}.}
\end{figure}

For simplicity, let us consider $N_1=N_2=1000000$ and that initially there are $200$ infectious individual in the non-confined population and zero in the isolated one. We show the exposed and infected fractions for isolated and non-confined individual in Fig. \ref{fig5} (a). As expected, the epidemic outbreak is more severe for non-confined individuals. However we despite the small disease transmission rate $\beta_{12}=\beta_{21}$, the total fraction of isolated persons that becomes sick is not negligible as can be seen in the accumulated number of sick people ($ I_i + R_i $) in Fig. \ref{fig5} (b).  

\section{Conclusions and future directions}
\lb{sec:C}
The proposed model of $n$ distinct populations interaction structured SEIR~\eqref{SEIR} stands as one alternative to describe the spread of COVID-19 pandemic diseases. The proposed model has the flexibility to include geographically separated communities as well as taking into account aging population groups and their interactions. Furthermore, we have shown how some strategies of non-pharmacological intervention as vertical and horizontal social isolation can be discussed using the proposed model.  
Analyzing some possible scenarios reflected by the proposed model, we show how a plateau-like curve of infected is a result of a disease's diffusion effect within distinct populations, resembling collected data from large countries as Brazil (see Figure~\ref{fig4}). In particular, we conjecture that the Covid-19 diseased diffusion from the capitals to the Brazil interior, as reflected by the proposed model, is responsible for plateau-like reported cases in the country. Collaborating with the analysis, we also present in Section~\ref{num} numerical solutions for some scenarios of the model~\eqref{SEIR} showing its applicability, in particular for describing the plateau-like shape of reported infected cases in Brazil.

\paragraph{Future directions:} 

Many open questions in regard to the proposed model~\eqref{SEIR} remains, and shall be explored in the future. Between them, we pointed out the following:

\begin{itemize}
    \item Investigate more realistic geometries, where a population interacts with more than two neighbors.
    \item Include more partition in the model, as for example dead and quarantine fractions. 
    \item Explore the complexity of the population interplay network more carefully, including age and distinct population interaction. 
    \item Analyze bifurcations and its consequences.
    \item Study optimal strategies for vaccination associated with the interacted population dynamics~\eqref{SEIR}. 
\end{itemize}

\end{document}